\documentclass[12pt]{article}

 \usepackage{paralist} 
     \usepackage{placeins} 
     \usepackage{amsmath,amsthm,amssymb}
     \usepackage{epsfig}
     \usepackage{psfrag}
     \usepackage{tabls}
     \usepackage{supertabular}
     \usepackage[square,comma,numbers]{natbib}
     \usepackage[top=2.5cm, bottom=2.5cm, left=2.5cm, right=2.5cm]{geometry}
\usepackage{indentfirst} 
\usepackage{fancyheadings}
\usepackage{colortbl} 
\definecolor{niceyellow}{rgb}{0.98,0.92,0.73}
    \newcommand{\myfecha}{\today}

\begin{document}
\begin{center}
\Large{\bf On the need to enhance physical insight via mathematical skills}
\end{center}
\begin{flushleft}
{\bf Sergio Rojas}\\
srojas@usb.ve \\
{Physics Department, Universidad Sim\'{o}n Bol\'{\i}var,
Ofic. 220, Apdo. 89000, Caracas 1080A, Venezuela.}
\end{flushleft}

{\bf Abstract}.
  It is becoming common to hear teaching advice about
spending more time on the ``physics of the problem"
so that students will get more physical insight
and develop a stronger intuition that can be very helpful
when thinking about physics problems.
Based on this type of justification, mathematical skills
such as the ability to compute moments of inertia, center of mass, or 
gravitational fields from mass distributions, and 
electrical fields from charge distributions are considered
``distracting mathematics" and therefore receive less attention. We argue 
a) that this approach can have a negative influence on
student reasoning when dealing with questions of rotational dynamics, a 
highly non-intuitive subject where even instructors may fail to provide 
correct answers, and b) that exposure of students to mathematical reasoning
and to a wide range of computational techniques 
to obtain the moment of inertia of different mass distributions will
make students more comfortable with the subject of rotational dynamics,
thus improving their physical insight on the topic.

\setlength{\parindent}{0.75cm}

\section{\label{sec:1}Introduction\protect}

  A common advice that the Nobel prize winning physicist
Lev Davidovich Landau often offered to students and colleagues 
approaching him about what and how to study, particularly to those 
interested in 
physics, was about the importance of mastering first the techniques of
working in the field of interest because ``fine points will come 
by itself." In his own words, ``You must start with mathematics which,
you know, is the foundation of our science.  [...] Bear in mind that 
by 'knowledge of mathematics' we
mean not just all kinds of theorems, but a practical ability to
integrate and to solve in quadratures ordinary differential equations, 
etc."
Or, in another response, ``What is needed is not all kinds of existence 
theorems, on which mathematicians lavish so much praise, but mathematical 
techniques, that is, the ability to solve concrete mathematical problems."
\cite{Lifshitz:1977}

  The importance of being able to express, interpret and manipulate
 physical results in mathematical
terms was also stressed by the great physicist Lord Kelvin ``I often say 
that when you can 
measure something and express it in numbers, you know something about it. 
When you can not measure it, when you can not express it in numbers, your 
knowledge is of a meager and unsatisfactory kind. It may be the beginning 
of knowledge, but you have scarcely in your thoughts advanced to the state 
of science, whatever it may be." \cite{Hewitt:1993} Freeman Dyson was
more eloquent ``...mathematics is not just a tool by means of which phenomena
can be calculated; it is the main source of concepts and principles by means of which new theories can be created."\cite{Dyson:1964}

The remarks of these great physicists bring to mind the challenges of 
undergraduate students learning rotational dynamics in the sciences and 
engineering fields while in a typical calculus-based physics course. 
Students face the problem of not finding sufficient detailed solved 
textbook examples of rotational inertia ($I$) computations. Despite 
the fact that in general ``intuition" cannot be used as a good guide 
for computing $I$
(although it can be of help when computing, for instance, the center of 
mass of a continuous body), no more than two or three worked examples 
can be found in most calculus-based physics 
textbooks\cite{HallidayResnickWalker:2000,SerwayJewett:2003,TiplerMosca:2003,note:1}\nocite{Rowley:2006}. 
Accordingly, based on published recent results\cite{RimoldiniSingh:2005} 
we would argue that students' 
limited exposure to or poor training 
in computing $I$ for several rigid body geometries with rotational
motion about different axis, can limit their
intuition\cite{Singh:2002} when asked to answer questions involving 
the comparison, for instance,
of the rotational kinetic energy of different objects under similar 
situations.

\section{\label{sec:2}About the Problem}

The difficulties that students have in realizing the dependence of $I$ on
the mass distribution about the rotating axis
is evident when they are asked
to compute the rotational inertia of a thin rod about one of the rod's 
ends and compare it with
the corresponding $I$ of a thin rectangular plate rotating about 
one of its edges.  Students become uneasy about the unexpected result
that in the latter case 
$I$ does not depend on the length of the edge about which the rotation 
takes place, and that $I$ has the same dependence on the length of 
the edges perpendicular to the rotational axis as the thin rod. 
Moreover, in spite of learning that the moment of inertia of a system of 
particles is the sum of the rotational inertia of individual particles
about the same rotational axis, students
are amused by the idea that to compute $I$ for an axis through the center
of mass and perpendicular
 to the plane of a rectangular thin plate of uniform mass $M$,
for instance, we can divide the object into four rectangular smaller plates,
each one of mass $M/4$ rotating around an axis through a corner
and perpendicular  to the surface of each lamina.
Then the desired $I$ is four times the rotational inertia of one of the
smaller plates.

Their anxiety diminishes somewhat when
they can compute $I$ using different methods 
such as direct integration, using the parallel axis theorem,
and dividing a complicated body into 
smaller pieces of known $I$ about the axis of interest.
They gain even more confidence about the correctness of the result after
finding the right answer in a textbook or after measuring a prediction
of a physical quantity that depends on $I$
(i.e. the period of oscillation of the corresponding physical pendulum
for both objects). Nevertheless, the experimental approach also requires
that students have first found $I$ mathematically. In other words,
it is required that students have
some conceptual knowledge before they carry out 
experiments\cite{AufschnaiterAufschnaiter:2007,note:2}.

Analogous ``surprising results" can be shown
to confuse students when doing similar computations of, for instance 
the moment of inertia of a hollow
thin cylinder and a solid cylinder, each one rotating around their 
respective axis. How can it be that $I$ for the latter case
is smaller than for the first case?

To be able to tackle the above exposed difficulties 
with some success,
students need to believe that these results are neither obvious nor intuitive.
Moreover, even for instructors many rotational dynamics outcomes 
are not intuitive\cite{Singh:2002}.  Accordingly, 
the old maxim ``{\em A repetitio studiorum mater est}" 
applies, but it has to be ``guided by critical feedback and {\em deliberate}
attempts to improve."\cite{Hestenes:2003} That is, the only way for 
students not to be surprised by surprising results is by doing  more
rotational inertia computations, which many colleagues dismiss as 
``{\em distracting mathematics}," which hardly help students gain  
physical insight.    

We might agree that in standard calculus courses students will learn,
among other mathematical skills, the formal techniques on how to
carry out some integrals. But it is in physics courses where students
start to apply what they have learned in their math classes and to
find new non-formal approaches to performing 
computations\cite{Yeatts:1992}. Some warnings on the traditional 
instruction of math in physics are also 
available\cite{RedisTalk:2005,Sherin:2001}.
 We believe that the
mathematical understanding of a problem is a process that involves 
meaningful learning which goes beyond the merely application of rote
procedures and involves ``higher order thinking skills."\cite{Rigden:1987}
Moreover, by using properly designed quantitative problems 
requiring students to illustrate their conceptual
learning and understanding will reveal much to the teachers and provide
invaluable feedback\cite{Reif:1981,ReifScott:1999,YapWong:2007,DunnBarbanel:2000}, and 
they can also be a powerful
means for helping students to understand the concepts of 
physics\cite{Rigden:1987,DunnBarbanel:2000}. 

\section{\label{sec:3}About the evidence}

In a study on students' understanding of rotational 
dynamic concepts, Rimoldini and Singh\cite{RimoldiniSingh:2005} point 
out some difficulties 
that students encounter when dealing with the subject. In particular, 
when students they interviewed were asked about
how the angular velocity
or the period of rotating objects would depend on the mass distribution
of the object
``Many did not use the concept of rotational inertia correctly. Some said
that they vaguely remember that the distribution of mass matters but did
not remember the exact relation."
This observation is a reflection of the
limited exposure students have
to the computation of the rotational inertia of different mass 
distributions and how it
actually depends on the mass distribution about the rotational axis.

This difficulty is further made explicit in a written test of thirty
multiple-choice questions that the authors\cite{RimoldiniSingh:2005} also 
administered to 652 
students from
calculus and algebra based introductory physics courses, which includes
an honor class of 97 students and an upper-level class of 17 physics majors
enrolled in an intermediate mechanics course. The test questions, a total
of thirty, are available as Appendix B of the
report\cite{RimoldiniSingh:2005}, and students were
required to provide justification for their answers.

In addition to two classes of 
student difficulties identified by Rimoldini and Singh,  a) those sharing 
a common ancestry with linear motion and b) those uniquely related to the 
more intricate nature of  rotational motion, we could add a third 
category c) those associated with insufficient training of students 
on the mathematical computation of  rotational inertia. 

Rimoldini and Singh found that some of the students they interviewed were 
uncertain about the meaning of $I$, which likely indicates that
the students had little practice in computing $I$ and had not mastered 
the techniques of computing $I$. Recalling that by mathematically
solving a problem involves a ``higher-order 
thinking skills,"\cite{Rigden:1987} 
we also share the idea expressed by
Rigden in the sense that ``a student's ability to discuss the 
problem--to do so in words of their own choosing, to do so clearly 
and accurately--indicates an understanding in which we can have confidence."
\cite{Rigden:1987}
And to get to that level some practice is required\cite{note:3}. 
In fact, in studies comparing conceptual learning and problem
solving skills, students enrolled in courses based on traditional instruction
scored on average higher in quantitative problems than students enrolled
in courses emphasizing conceptual 
learning\cite{HoellwarthEtAll:2005,Lea:1994,note:4}.
As pointed out by L. D. Landau ``I can only 
emphasize that you must perform
all the calculations by yourself, and must not leave it to the authors of
the books you have read."\cite{Lifshitz:1977} Though the ability to do 
so is a matter of training in ``distracting mathematical computations" 
such as computing rotational inertia, the training should also
involve a proper physical interpretation of the obtained
quantitative results\cite{note:5,ReifScott:1999}.

To make our third category more evident, we first concentrate
on the responses to the thirty questions given by students of
the honor class. The answers can be divided into  two
sets: those that were answered correctly by eighty or more percent of
the students (high level of correct response) and those that
were answered correctly by less than eighty percent of the students.
We find 14 questions in the first group. In the second group
fall the remaining 16 questions. The answers to the questions are given
on page 7 of the report\cite{RimoldiniSingh:2005}.

It is expected that students of an honor class would be able to perform 
standard computations involving integration of not too complicated expressions.
 Correspondingly, according  to this expectation we would expect that
all 7 of the questions involving knowledge of the rotational
inertia of the rotating solid (i. e. questions 1, 3, 4, 20, 24, 25, and 29)
would have received a higher level of correct response (to make the
article self-contained we are including these 7 questions as appendix A and
the respective responses to these questions are shown in Table~\ref{table1}). 
Such an expectation is reasonable because the required computations are 
fairly easy.  The results show that only question number 20
was answered in conformity with this expectation, and it was answered 
correctly by 85 percent of the students of the honor class
(because the computation requires a simple
integration and, more importantly, the rotational inertia of
a homogeneous cylinder is one of the few textbook worked out examples,
this level of response should have been much better).
By similar reasoning, questions
1, 3, and 4 should have received a higher level of correct responses
by the honor class.

We believe that  if students of the honor class had received enough training
in computing the rotational inertia
of composite and simple objects (like the ones involved in these
questions), they
would have given a higher level of correct response on 7 of the 
questions requiring these simple 
computations\cite{ReifScott:1999,Rigden:1987,Lea:1994,YapWong:2007,note:5}.
This would have reduced the gap among the questions
receiving a high level of correct responses and those that did not.

These observations also apply to the other group of students that took the
test. For example, while 76 percent of honors students provided the correct 
response to question 3 (see Table~\ref{table1}),
physics majors in an upper-level class did not
perform better on this than non-honors introductory students (41 and 45 
percent correct responses respectively).
The upper level class did better in answering question 4 than the 
other two. Regarding rotational inertia,
these were the only two questions common to each group.
On average, 71 percent of the students from the honor class answered correctly
all the questions regarding rotational inertia while 56 percent of
the non-honor
class answered correctly the same questions. 

We feel that since the
mathematical computations and algebraic manipulations involved in these 
questions are not really demanding, the failure of honor students to 
respond correctly is a consequence of neglecting mathematical skills over 
physical 
insight\cite{Hestenes:2003,MualemEylon:2007,Sherin:2001,Yeatts:1992}.
In addition, in their 
study Hoellwarth {\em et. al.}\cite{HoellwarthEtAll:2005}
conclude that ``students must be taught both concepts and problem solving
 skills explicitly if we want students to be proficient at both." In this
sense, proactive teaching strategies should lead to the identification
of quantitative problems helpful to recognize both conceptual and
quantitative understanding of students. In fact, 
some fruitful ideas have been advanced on how to properly
address the design of instruction so the involved learning cognitive mechanism
of the students are triggered, leading to a more effective teaching
outcomes\cite{Hestenes:2003,DunnBarbanel:2000,Reif:1981,ReifScott:1999,RedishSteinberg:1999,Scherr:2007}.

\section{Concluding Remarks}

We propose that the common difficulty students have in answering 
correctly questions of physical quantities involving rotational inertia 
is likely rooted in the limited exposure students have 
to computing and analyzing these quantities because these techniques, while 
essential, are considered  to be ``distracting mathematics,"  and their 
importance is not emphasized by 
instructors\cite{Hestenes:2003,Reif:1981,ReifScott:1999}. Another 
reasons for students' difficulty 
is that textbooks used by students provide just one or two simple examples 
as models for students to learn these computational 
techniques\cite{note:1}.
Considering the non-intuitiveness of rotational dynamics, even for
instructors having  wide experience teaching the subject\cite{Singh:2002},
 both reasons
conspire against students reasoning on this subject.

In the analysis of the collected data, on page 6 of the 
report\cite{RimoldiniSingh:2005} Rimoldini and Singh pointed out that
``many students were unsure about this concept. For example, many did
not know that $I$ is a function of the mass distribution about an axis
and that the rotational kinetic energy depends on $I$ and not just
on the total mass.[...] Interviews showed that this type of difficulty
was partly due to the students' unfamiliarity with $I$".

In some sense the research of Rimoldini and Singh somehow supports
the idea that because of an 
overemphasis\cite{MualemEylon:2007,HoellwarthEtAll:2005,note:6} on
the qualitative (conceptual) physical aspects of the problems, 
standard mathematical abilities, which are essential for understanding 
the whole physical process are not taught because, rephrasing a passage 
from a recent editorial\cite{Klein:2007}, they
interfere with the students' emerging sense of physical insight. 

Thus, if instructors do not have enough time to train students relevant 
computational techniques, textbook publishers  should not leave mathematical 
computations only 
to the students. In addition to rotational inertia, textbooks should also 
include more solved illustrative examples 
on computing center of mass, gravitational and electric fields\cite{note:1}
and constantly point out that the involved techniques  
are essentially the same\cite{Yeatts:1992}.
This is an important requirement for a textbook because innovative 
active-learning teaching strategies requires students to acquire basic and 
fundamental knowledge through reading a textbook.
Certainly, innovative teaching strategies will help
in handling thicker and
heavier textbooks, with lots of physical and mathematical insights within
them\cite{YapWong:2007,HoellwarthEtAll:2005,CrouchMazur:2001,James:2006,CerbinKopp:2006}.

To paraphrase Heron and Meltzer,
learning to approach problems in a systematic way starts from teaching
and learning the interrelationships among conceptual knowledge, 
mathematical 
skills and logical reasoning\cite{HeronMeltzer:2005}.
In physics, this necessarily requires the teaching of a good 
deal of ``distracting mathematical computations."

\appendix
\section{Test problems}
To make the article self-contained, in this section we are including 
the 7 multiple choice test problems we are 
analyzing\cite{RimoldiniSingh:2005}.

{\bf 1}. Two copper disks (labeled ``A" and ``B") have the same radius but 
disk B is thicker with four times the mass of disk A. They spin on 
frictionless axles. If disk A is rotating twice as fast as disk B, which 
disk has more rotational kinetic energy? \\
(a) The faster rotating disk A. 
(b) The thicker disk B. 
(c) Both disks have the same rotational kinetic energy. 
(d) It depends on the actual numerical values of the angular speeds 
    of the disks. 
(e) None of the above. 

{\bf 3}. An aluminum disk and an iron wheel (with spokes of negligible mass) 
   have 
   the same mass M and radius R.  They are spinning around their frictionless 
   axles with the same angular speed as shown. Which of them has 
   more rotational kinetic energy? \\
(a) The aluminum disk. 
(b) The iron wheel. 
(c) Both have the same rotational kinetic energy. 
(d) It depends on the actual numerical value of the mass M. 
(e) None of the above. 

{\bf 4}. Consider the moment of inertia, I, of the rigid homogeneous disk of 
   mass M shown below, about an axis through its center (different shadings 
   only differentiate the two parts of the disk, each with equal mass M/2). 
   Which one of the following statements concerning I is correct? \\
(a) The inner and outer parts of the disk, each with mass M/2 (see figure), 
    contribute equal amounts to I. 
(b) The inner part of the disk contributes more to I than the outer part. 
(c) The inner part of the disk contributes less to I than the outer part. 
(d) The inner part of the disk may contribute more or less to I than the 
    outer part depending on the actual numerical value of the mass M of 
    the disk. 
(e) None of the above. 

{\bf 20}. The moment of inertia of a rigid cylinder \\
(a) does not depend on the radius of the cylinder. 
(b) does not depend on the mass of the cylinder. 
(c) depends on the choice of rotation axis. 
(d) depends on the angular acceleration of the cylinder. 
(e) can be expressed in units of kg. 

{\underline{Setup for the next three questions}} 
   An aluminum disk and an iron wheel (with spokes of negligible mass) have 
    the same radius R and mass M as shown below. Each is free to rotate about 
    its own fixed horizontal frictionless axle. Both objects are initially at 
    rest. {\underline{Identical}} small lumps of clay are attached to 
    their rims as shown in 
    the figure (the figure shows each rim on vertical position and
    the small mass attached to the right of the rim on the horizontal
    diameter). 

{\bf 24}. Which one of the following statements about their angular 
     accelerations is true? \\ 
(a) The angular acceleration is greater for the disk+clay system. 
(b) The angular acceleration is greater for the wheel+clay system. 
(c) Which system has a greater angular acceleration depends on the actual 
    numerical values of R and M. 
(d) There is no angular acceleration for either system. 
(e) The angular accelerations of both systems are equal and non-zero. 

{\bf 25}. Which one of the following statements about their maximum angular 
    velocities is true? \\ 
(a) The maximum angular velocity is greater for the disk+clay system. 
(b) The maximum angular velocity is greater for the wheel+clay system. 
(c) Which object has a greater maximum angular velocity is determined by 
    the actual numerical values of R and M. 
(d) The maximum angular velocities of both systems are equal and non-zero. 
(e) There is no angular velocity for either system so the question of a 
    maximum value does not arise. 

{\underline{Setup for the next two questions}} 
Two copper disks of different thicknesses have the same radius but 
different masses as shown below. Each disk is free to rotate about its own 
fixed horizontal frictionless axle. Both disks are initially at rest. 
{\underline{Identical}} small lumps of clay are attached to their rims 
as shown in the figure. (the figure shows each rim on vertical position and
    the small mass attached to the right of the rim on the horizontal
    diameter). 

{\bf 29}. Which one of the following statements about their angular 
    accelerations is true? \\ 
(a) The angular acceleration is greater for the system in which the disk 
    has larger mass. 
(b) The angular acceleration is greater for the system in which the disk 
    has smaller mass. 
(c) Which system has a greater angular acceleration depends on the actual 
    numerical values of their masses. 
(d) There is no angular acceleration for either system. 
(e) The angular accelerations of both systems are equal and non-zero. 

\section*{Acknowledgments}
  We are grateful to
  Dr. Cheryl Pahaham, who kindly provided 
useful comments 
  on improving this article.

\bibliographystyle{unsrt}
\bibliography{RojasS_submission_ref} 

\begin{table*}[h]
\begin{center}
\caption{\label{table1}
Multiple-choice questions (included in Appendix A) were administered to a 
total of 669 students. The performance of 559 general (calculus- and 
algebra-based) introductory non-honor students (GI) is distinguished from 
an honor class (HC) of 93 introductory students, and an upper-level 
(UL) class of 17 physics majors enrolled in an intermediate mechanics course 
(who were administered a subset of 11 questions).  The table presents the 
average percentage (rounded to the nearest integer) of students selecting 
the answer choices (a)--(e) for each question of the test 
(bold numbers refer to the correct responses).
}
\begin{tabular}{|cccccccccccccccc|}
\hline
Answers   &   &(a)&   &   &(b)&   &   &(c)&   &   &(d)&   &   &(e)&  \\
\hline
Questions &GI &HC &UL &GI &HC &UL &GI &HC &UL &GI &HC &UL &GI &HC &UL \\
\hline
    1     &16 &10 &   &23 &16 &   &{\bf 57} &{\bf 73} &   &4  &1  &   &1  &0  &   \\
    3     &18 &27 &29 &{\bf 45} &{\bf 71} &{\bf 41} &35 &2  &18 &1  &0  &6  &1  &0  &6 \\
    4     &13 &7  &6  &22 &15 &6  &{\bf 61} &{\bf 76} &{\bf 82} &3  &0  &6  &0  &2  &0 \\
   20     &2  &2  &   &3  &2  &   &{\bf 71} &{\bf 85} &   &18 &9   &   &6  &2  & \\
   24     &{\bf 33} &{\bf 61} &   &32 &24 &   &3  &0  &   &4  &0   &   &28 &15 & \\
   25     &{\bf 28} &{\bf 58} &   &32 &21 &   &4  &2  &   &34 &18  &   &2  &1  & \\
   29     &19 &9  &   &{\bf 60} &{\bf 75} &   &3  &0  &   &1  &1   &   &17 &15 & \\
\hline
\end{tabular}
\end{center}
\end{table*}
\label{LastPage}
\end{document}